Title: On the origin of zombies: a modeling approach

Authors: Alisha Kumari[1#], Elijah Reece[2*#], Kursad Tosun[3], Scott Greenhalgh[3]

1. Department of Physics, Siena College, 515 Loudon Road, Loudonville, NY, 12211
2. Department of Biology, Siena College, 515 Loudon Road, Loudonville, NY, 12211
3. Department of Mathematics, Siena College, 515 Loudon Road, Loudonville, NY, 12211

*-corresponding author
#-authors contributed equally

**Abstract**

A zombie apocalypse is one pandemic that would likely be worse than anything humanity has ever seen. However, despite the mechanisms for zombie uprisings in pop culture, it is unknown whether zombies, from an evolutionary point of view, can actually rise from the dead. To provide insight into this unknown, we created a mathematical model that predicts the trajectory of human and zombie populations during a zombie apocalypse. We parameterized our model according to the demographics of the US, the zombie literature, and then conducted an evolutionary invasion analysis to determine conditions that permit the evolution of zombies. Our results indicate a zombie invasion is theoretically possible, provided there is a sufficiently large ratio of transmission rate to the zombie death rate. While achieving this ratio is uncommon in nature, the existence of zombie ant fungus illustrates it is possible and thereby suggests that a zombie apocalypse among humans could occur.

# 1. Introduction

The world is continuously at risk from pandemics, with Covid-19, SARS, and Ebola serving as recent examples of their devastating impacts. As time progresses, diseases capable of starting another pandemic are more than likely to occur. One important potential pandemic that would likely be worse than anything humanity has ever seen is a zombie apocalypse. While this may seem far-fetched for humanity, in South America among other regions, a fungus exists that can turn ants into zombie ants (*Zombie Ant*, 2021), which implies such an outbreak among humans is within the realm of biological possibilities.

What is biological possible constitutes all species, most of which exhibit enormous diversity of traits ("Coevolution of Habitat Diversity and Species Diversity," 1995; Nowak, 2006). Through examining these traits, specifically the trade-offs between them (Nowak, 2006), the direction of evolution can be inferred, which can provide a glimpse as to what may be in store for a species' future. Typically, such an evolution is caused by the occurrence of a rare mutant, or a patient zero in the case of a novel disease (McKay, 2017; Peters, 2014), which features some form of trait advantage in reproductive ability, size, speed, susceptibility to disease, or survival rate, among others.

While patient zero is ubiquitous in many pop culture movies and tv shows as the first individual to become a zombie (Wikipedia contributors, 2021) the mechanism by which the first zombie is created is often relatively unknown. Classically, many possible scenarios lead to the uprising of patient zero, and ultimately a full-blown zombie apocalypse. For instance, consumption of the mutated zombie ant fungi could infect humans, causing them to seek out nutrients by cannibalism, and thereby further spread fungal spores through their saliva (*The Last of Us*, n.d.). Alternatively, medical experimentation is often a culprit in causing patient zero, with cross-transmission events from monkeys to humans (Boyle & Garland, 2003; Fresnadillo & Joffe, 2007), and side-effects of untested vaccines (Lawrence et al., 2007) standing as common causes. However, despite these mechanisms for zombie uprisings, it is unknown whether zombies, from an evolutionary point of view, can actually rise from the dead. So, to provide insight into this unknown we created a mathematical model that predicts the trajectory of human and zombie populations during a zombie apocalypse. Using this mathematical model, we apply stability analysis (Martcheva, 2015) to estimate the long term prognosis of the United States, conduct an evolutionary invasion analysis (Otto & Day, 2011) to infer conditions that allow the zombie apocalypse to occur, or invade from another country, and investigate the potential for an endless human-zombie war through Hopf bifurcation analysis. Our main findings show an uprising of zombies requires the fungus to transmit their spores to more than $0.023$ *humans per day,* and would likely lead to an oscillating struggle that decreases over thousands of years between humans and zombies, as both try to overwhelm the other.

## 2. Methods

To determine the conditions that permit the evolution of zombies, we developed a mathematical model of zombie transmission in a human population. We calibrate our model to the demographics of the US and then apply stability, evolutionary invasion, and Hopf-bifurcation analyses to inform on the long-term outcomes for humanity.

### 2.1 Mathematical model

To begin, we created a mathematical model that predicts the long-term population of the US. We then extend the model to include zombies and proceed to investigate the model's long-term behavior.

2.1.1. The resident system. We first consider a resident system of humans split into two compartments. One compartment represents the population of humans in the United States ($N$), which is the population of susceptible humans, and the second represents the number of deceased humans due to natural causes, which have yet to completely decompose ($D$). The rates governing the transition between these compartments is given by

$$\frac{dN}{dt} = b(1 - \frac{N}{K})N - \delta N, \qquad (1)$$

$$\frac{dD}{dt} = \delta N - \mu D,$$

where $b$ is the birth rate, $\delta$ is the mortality rate of people in the US during the year 2020, $K$ is the limiting capacity of humans in the US, and $\mu$ is the rate at which dead bodies decompose.

Table 1. Parameters, base values, and sources.

| Constant | Parameter | Value | Citation |
| --- | --- | --- | --- |
| $\beta(\delta_m)$ | Transmission rate | 0.35 per day | (Pacheco, J., 2021) |
| $b$ | Birth rate | 0.000065753 per day | ("United States," 1968) |
| $\delta$ | Mortality rate | 0.00003562 per day | ("United States," 1968) |
| $\mu$ | Decomposition rate of a human body | 0.00595 per day | (Adams, 2017) |
| $K$ | Population capacity | 4,672,507,360 people | Section 2.2 |
| $\gamma$ | Zombie death rate | 0.01 corpses per day | (archyde, 2021) |

2.1.2. The zombie equation. We also consider a third compartment that tracks the number of humans that have been turned into zombies ($Z$). This compartment is governed by the differential equation,

$$\frac{dZ}{dt} = \frac{\beta}{K} NZ - \gamma Z \qquad (2)$$

where $\gamma$ is the zombie death rate, and $\beta$ is the transmission rate of zombism.

2.1.2. The extended system. The extended system is a combination of the resident system and the zombie equation. The equations are linked by including a transmission rate to capture the spread of zombism and a mortality rate that reflects patient zero naturally rising from the dead. Altogether, this yields:

$$\begin{aligned}\frac{dN}{dt} &= b(1 - \frac{N}{K})N - \delta N - \frac{\beta(\delta)}{K} D N - \delta_M N - \frac{\beta(\delta_M)}{K} Z N, \\ \frac{dD}{dt} &= (\delta + \frac{\beta(\delta)}{K} D)N - \mu D, \\ \frac{dZ}{dt} &= (\delta_M + \frac{\beta(\delta_M)}{K} Z)N - \gamma Z.\end{aligned} \qquad (3)$$

Note, the transmission term, $\beta$ is assumed to be a function of the death rate, $\delta$. When $\delta = 1/77/365$ that $\beta(\delta) = 0$ because natural death isn't transmittable.

2.2 Parameter estimation

We estimated the population capacity, $K$, from publically available data, and determined the value of transmission rate, $\beta$, based on the spread of zombie outbreaks from the literature (Pacheco, J., 2021). In addition, we also obtained $\gamma$ from the literature (archyde, 2021)

2.2.1 Population capacity

To estimate the population capacity $K$ In the US, we used a least-squares method. This method used the population of the US from 1960, 1980, 2000, and 2020 (*Population, Total - United States*, n.d.) (Figure 1), using the average lifespan of a person within the US, 79 years (Wamsley, 2021), in conjunction with predictions of the US population from the resident model (Figure 2). Through this procedure, the $K$ value that had the least-squares error was $K = 558,075,379$.

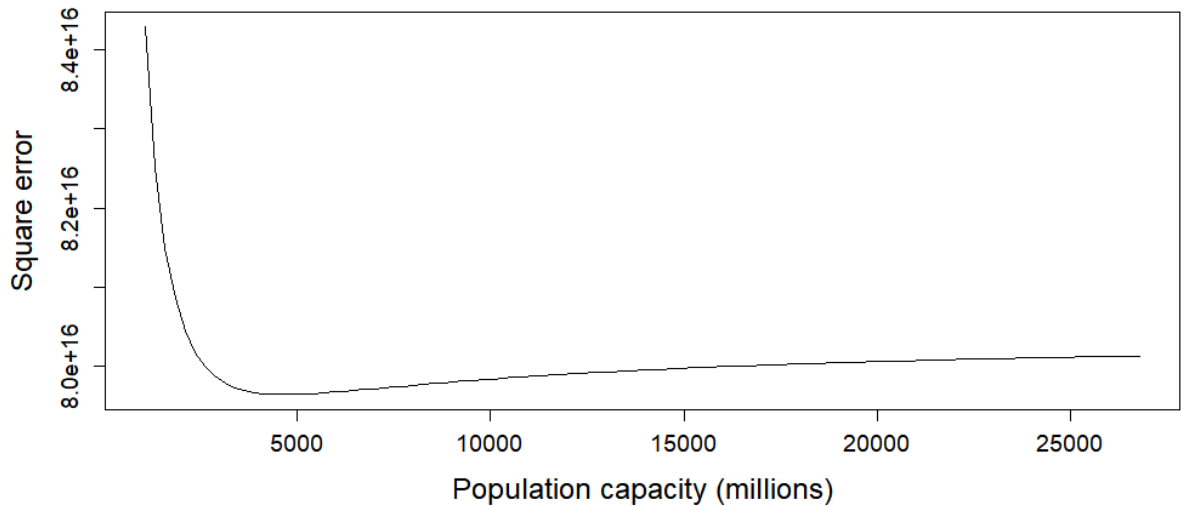

Figure 1. Square error of resident model from United States population. The square error between the predictions of the resident model and US population data for the given value of population capacity.

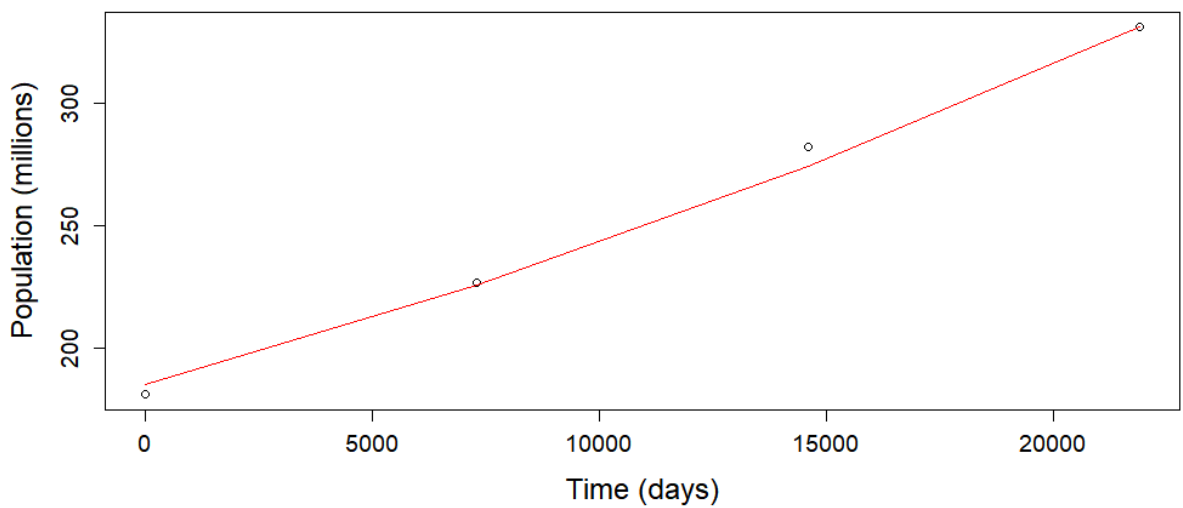

Figure 2. Resident model vs US population. The resident model with $K = 558,075,379$ (red line) and the US population from 1960 to 2020 (black points).

2.3 Stability analysis

Here, we apply stability analysis to the resident system to determine its long-term behavior. So, we evaluated the Jacobian at the non-extinction and extinction equilibria to determine the long-term behavior of the resident system, as characterized by its eigenvalues.

The Jacobian of system (1) at the non-extinction equilibrium, $\hat{N} = (1 - \frac{\delta}{b})K$ and $\hat{D} = \frac{\delta}{\mu}(1 - \frac{\delta}{b})K$, is

$$J_{res}\Big|_{\hat{N}=(1-\frac{\delta}{b})K,\ \hat{D}=\frac{\delta}{\mu}(1-\frac{\delta}{b})K} = \begin{pmatrix} -b+\delta & 0 \\ \delta & -\mu \end{pmatrix}$$

which yields the eigenvalues of $\lambda_1 = -b + \delta$, and $\lambda_2 = -\mu$. Thus, the non-extension equilibrium is locally stable provided the death rate is lower than the birth rate, $\delta < b$.

For the extinction equilibrium, $\tilde{N} = 0$ and $\tilde{D} = 0$, the Jacobian of the system (1) simplifies to

$$J_{res}\Big|_{\tilde{N}=0,\tilde{D}=0} = \begin{pmatrix} b-\delta & 0 \\ \delta & -\mu \end{pmatrix}.$$

The associated eigenvalues are $\lambda_1 = b - \delta$, and $\lambda_2 = -\mu$. The extinction equilibrium is thus locally stable when the death rate is greater than the birth rate, $\delta > b$.

2.4 Analysis of zombie invasion and evolution

To determine whether zombies could evolve and become the dominant form of death, we extend the resident model to include a Zombie class, $Z$. We then provide details of the Jacobian of the extended model to illustrate conditions that permit zombies to invade a population and their potential evolution from the dead.

2.4.1. Conditions for a zombie invasion. For the extended system (3), when $\beta(\delta) = 0$ and $\delta_m = 0$, the Jacobian at $\hat{N} = (1 - \frac{\delta}{b})K, \hat{D} = \frac{\delta}{\mu}(1 - \frac{\delta}{b})K$ and $\hat{Z} = 0$ is

$$J\Big|_{N=(1-\frac{\delta}{b})K,\ D=\frac{\delta}{\mu}(1-\frac{\delta}{b}),\ Z=0} = \begin{pmatrix} -b+\delta & 0 & -\beta(\delta_m)\left(1-\frac{\delta}{b}\right) \\ \delta & -\mu & 0 \\ 0 & 0 & \beta(\delta_m)\left(1-\frac{\delta}{b}\right)-\gamma \end{pmatrix}.$$

Thus, given $\lambda_1 = -b + \delta < 0$, $\lambda_2 = -\mu < 0$, it follow that zombies cannot invade provide

$$\lambda_3 = \beta(\delta_m)(1 - \frac{\delta}{b}) - \gamma < 0$$

2.4.2. Conditions that prevent the zombie uprising. To examine the potential uprising of patient zero, we now consider the Jacobian of the extended system for $\delta_m$ close, but not equal, to zero. Specifically, the Jacobian evaluated at $\hat{N} = (1 - \frac{\delta}{b})K$, $\hat{D} = \frac{\delta}{\mu}(1 - \frac{\delta}{b})K$, and $\hat{Z} = 0$ is

$$J_{ext}\Big|_{N=\widehat{N}, D=\widehat{D}, Z=0} = \begin{pmatrix} b\left(1 - \dfrac{2\widehat{N}}{K}\right) - \delta - \delta_m & 0 & -\dfrac{\beta(\delta_m)\widehat{N}}{K} \\ \delta & -\mu & 0 \\ \delta_m & 0 & \dfrac{\beta(\delta_m)\widehat{N}}{K} - \gamma \end{pmatrix}.$$

It follows that the eigenvalues are:

$$\lambda_1(\delta, \delta_m) = -\mu,$$

$$\lambda_2(\delta, \delta_m) = \tfrac{1}{2}\left((1 - \tfrac{\delta}{b})\beta(\delta_m) - b + \delta - \delta_m - \gamma \right.$$
$$+ \sqrt{(1 - \tfrac{\delta}{b})\beta(\delta_m)^2 + 2(1 - \tfrac{\delta}{b})(b - \gamma - \delta_m - \delta)\beta(\delta_m) + (\delta_m - \delta - \gamma + b)^2}\,),$$

$$\lambda_3(\delta, \delta_m) = \tfrac{1}{2}\left((1 - \tfrac{\delta}{b})\beta(\delta_m) - b + \delta - \delta_m - \gamma \right.$$
$$- \sqrt{(1 - \tfrac{\delta}{b})\beta(\delta_m)^2 + 2(1 - \tfrac{\delta}{b})(b - \gamma - \delta_m - \delta)\beta(\delta_m) + (\delta_m - \delta - \gamma + b)^2}\,).$$

Thus, for the zombie-free equilibrium to be an evolutionarily stable state (Otto & Day, 2011), we require

$$\lambda_2(\delta^*, \delta_m^*) \geq \lambda_2(\delta^*, \delta_m),$$

where $\delta^* = 1/77/365$, and $\delta_m^* \approx 0$.

For values of $\delta_m$ close to 0, we have that

$$\lambda_2(\delta^*, \delta_m) \approx \lambda_2(\delta, 0) + \dfrac{\partial \lambda_2(\delta, 0)}{\partial \delta_m}(\delta_m - 0),$$

where $\lambda_2(\delta^*, 0) = -\gamma$, and $\dfrac{\partial \lambda_2(\delta^*, \delta_m)}{\partial \delta_m}\Big|_{\delta_m=0} = \dfrac{d\beta(0)}{d\delta_m}(1 - \dfrac{\delta^*}{b})$.

Therefore, for $\delta_m$ close to 0, the zombie-free equilibrium is an evolutionarily stable state-provided

$$\dfrac{\partial \lambda_2(\delta^*, \delta_m)}{\partial \delta_m}\Big|_{\delta_m=0} = 0 \Leftrightarrow \dfrac{d\beta(0)}{d\delta_m} = 0,$$

and

$$\dfrac{\partial^2 \lambda_2(\delta^*, \delta_m)}{\partial^2 \delta_m}\Big|_{\delta_m=0} < 0.$$

2.5 Periodic behavior

We now examine the potential for periodic behavior in the dynamics between humans and zombies by means of Hopf bifurcation analysis.

To begin, we assume $\delta_m \approx 0$. Thus, the extended system has the non-extinction and zombie endemic equilibria:

$$\bar{N} = \frac{\gamma}{\beta}K, \qquad \bar{D} = \frac{\gamma}{\beta}\frac{\delta}{\mu}K, \qquad \bar{Z} = \frac{((b-\delta)\beta - \gamma b)}{\beta^2}K$$

Rearranging the order of the system, computing the Jacobian and evaluating it at the non-extinction and zombie endemic equilibrium, we have that

$$\hat{J}\Big|_{N=\bar{N}, D=\bar{D}, Z=\bar{Z}} = \begin{pmatrix} -\mu & \delta & 0 \\ 0 & b\left(1 - \frac{2\bar{N}}{K}\right) - \delta - \frac{\beta}{K}\bar{Z} & -\frac{\beta \bar{N}}{K} \\ 0 & \frac{\beta}{K}\bar{Z} & \frac{\beta \bar{N}}{K} - \gamma \end{pmatrix}.$$

It follows that the eigenvalues of $\hat{J}\Big|_{N=\hat{N}, D=\hat{D}, Z=\hat{Z}}$ are

$$\lambda_1 = -\mu \text{ and } \lambda_{2,3} = \frac{1}{2}\left(-\frac{\gamma b}{\beta} \pm \sqrt{\left(\frac{\gamma b}{\beta}\right)^2 - 4\gamma\left(b - \delta - \frac{\gamma b}{\beta}\right)}\right).$$

For periodic behavior to occur we require purely imaginary eigenvalues, and so $\frac{\gamma b}{\beta} = 0$. If $\gamma = 0$ then $\lambda_{2,3} = 0$, which implies periodic behavior does not occur. If $b = 0$ then $\lambda_{2,3} = \pm \sqrt{\gamma \delta}$. Thus, for $\delta \geq 0$ and $\gamma \geq 0$ periodic behavior does not occur.

### 3. Results

To illustrate our predictions on the likelihood of a zombie apocalypse, and its effect on human populations, we parameterized our model according to the demographics of the US, and the zombie literature. Furthermore, to illustrate the potential outcomes for humanity and zombies, we evaluate the trajectory of our model for $\beta = 0.015$, $\beta = 0.029$, and $\beta = 0.35$, and determined the long-term behavior of the model through stability, evolutionary invasion analysis, and Hopf-bifurcation analyses.

In the absence of zombies, the US population converges towards maximum capacity, as nothing is hindering their growth. When zombies are included, the behavior of the system depends critically on the values of $\beta$ and $\delta_m$. For instance, given $\delta_m \approx 0$, the value of $\beta$ must be greater than 0.023 for zombies to disrupt the stability of the non-extinction equilibrium (Figure 5). Similarly, when $\beta = 0.35$ it is required that $\delta_m < 0.1575$ for zombies to be able to disrupt the stability of the non-extinction equilibrium (Figure 4).

The phases of the extended model show the pattern the outbreak could take, depending on how fast or slow zombies spread (Figure 7). For the zombie apocalypse to occur, the β value must be greater than 0.023. A value of β less than 0.023 causes the non-extinction and zombie endemic equilibrium to be unstable. For example, with a low value of β, such as 0.015, the zombies die out and the humans converge to their carrying capacity, K (Figure 6A, D, G). When β is slightly above 0.023, for example, 0.029, the system approaches a non-extinction and zombie endemic equilibrium, implying zombie and human populations end up coexisting (Figure 6B, E, H). For higher values of β, such as 0.35, shows more frequent decreasing oscillations between human and zombie populations, implying both populations will battle it out for dominance (Figures 6 C, F, I, and Figure 7).

To determine if the current form of death is an evolutionarily stable state, we examine the largest eigenvalue $\lambda_2(\delta^*, \delta_m^*)$ when $\delta_m^* \approx 0$ (Figure 4). Specifically, for $\delta_m > \delta_m^*$, we have that $\lambda_2(\delta^*, \delta_m^*) > \lambda_2(\delta^*, \delta_m)$ (Figure 4). This means that natural death without zombies is the dominant form of death for humans, which implies that zombies cannot evolve from the dead.

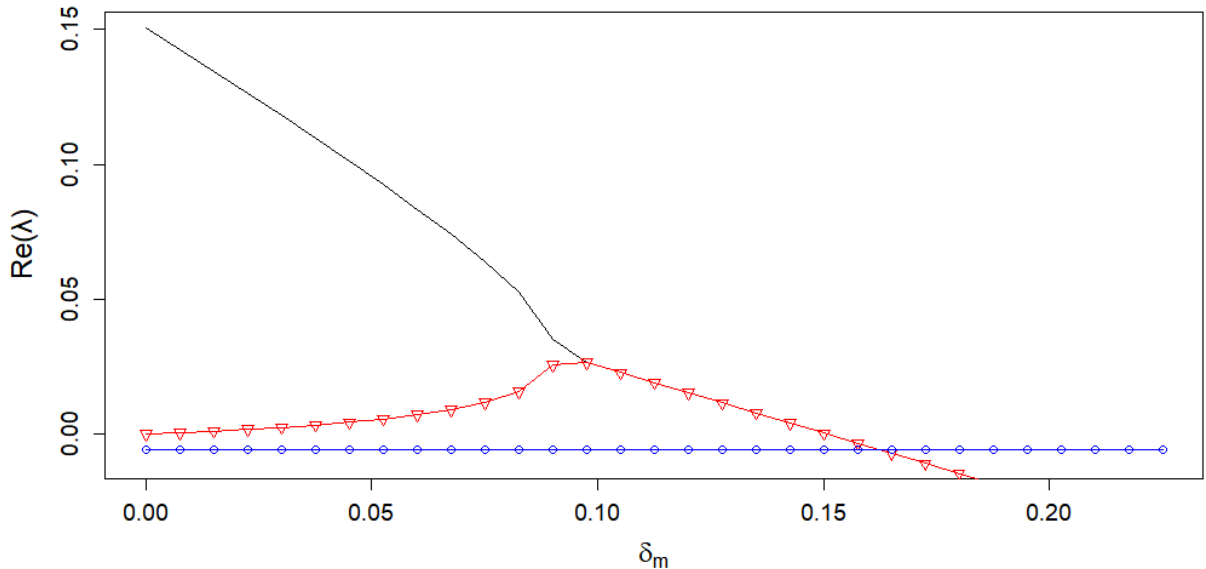

Figure 4. The change in the real part of the eigenvalues of the extended system with respect too $\delta_m$. The blue line with circles is the eigenvalue $-\mu$, while the red line with triangles and black line represents the real parts of the eigenvalues $\lambda_{2,3}$, respectively. When $Re(\lambda) > 0$ for any eigenvalue a zombie outbreak can occur.

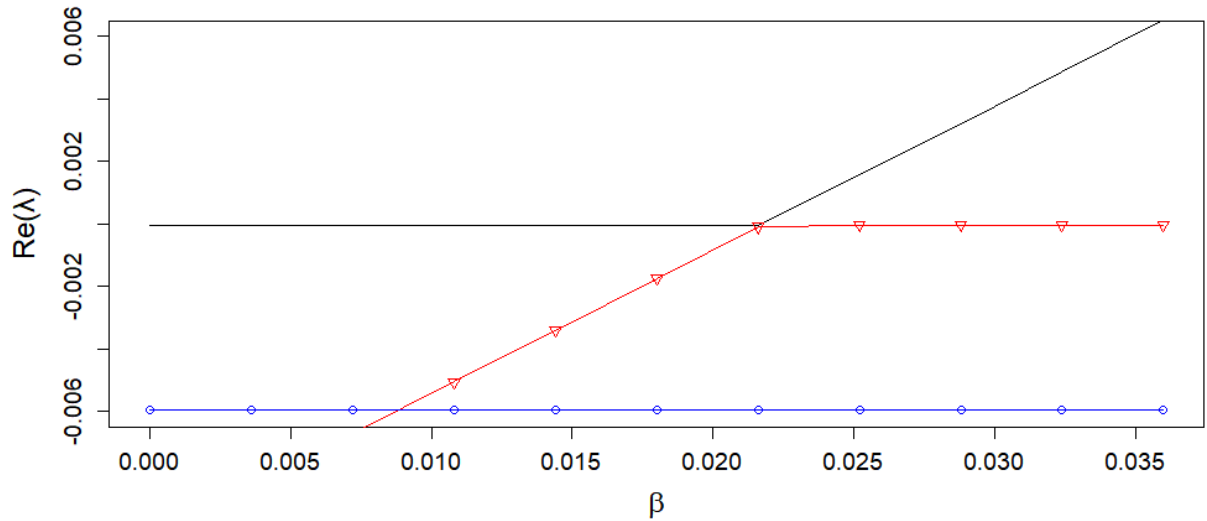

Figure 5. The change in eigenvalues of the extended system with respect to β. The value of $\lambda_1 = -\mu$ is shown by the blue line with circles. The real part is represented by the red line with triangles and black lines, respectively. The critical point on this graph is where $\lambda_2$ is greater than 0, which occurs when $\beta \approx 0.023 \; per \; day$. Values of result in a zombie outbreak.

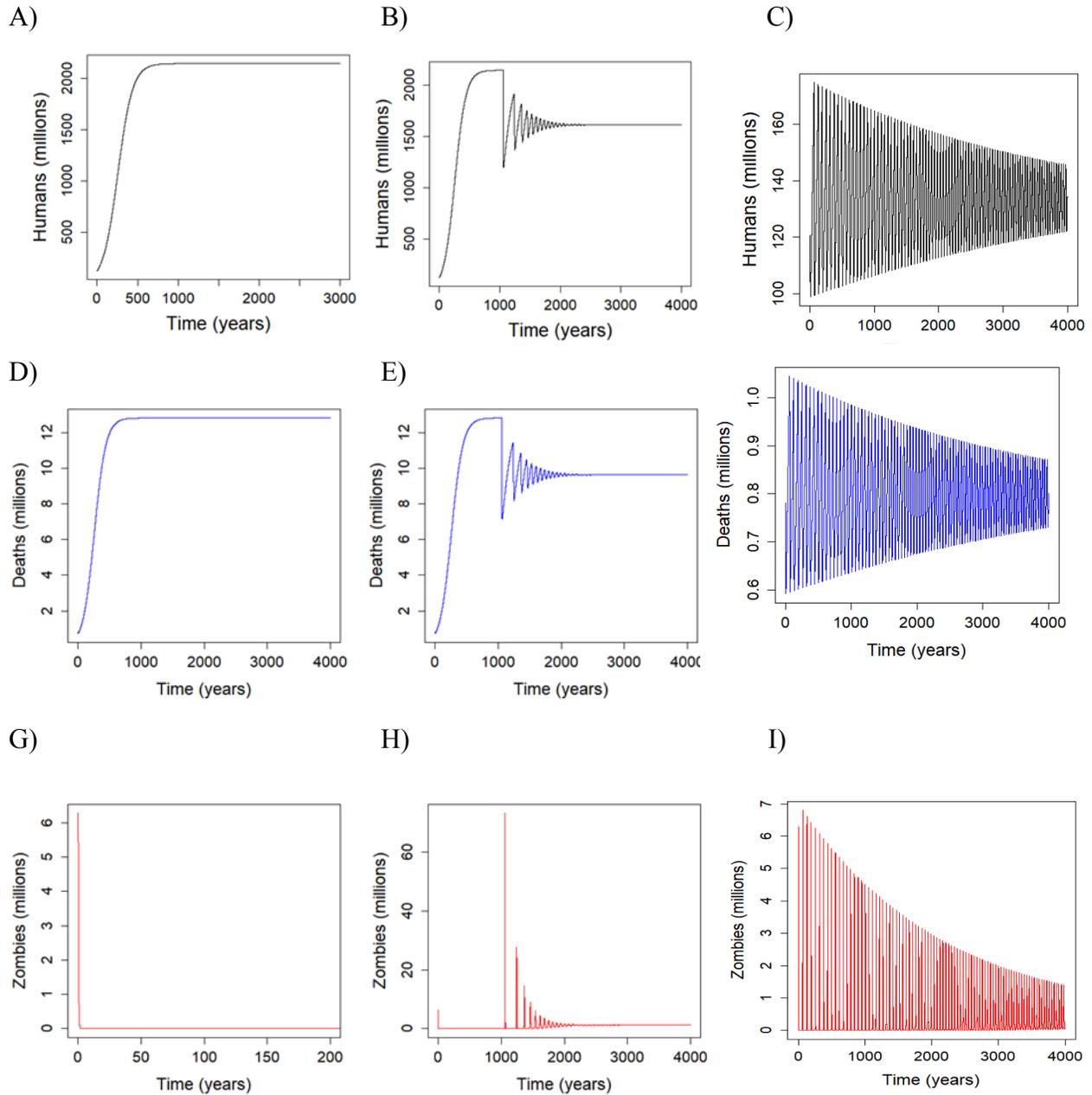

Figure 6. 3 The trajectory of the extended model. From left to right, the columns have β values of 0.015, 0.029, and 0.35. The first column shows what happens when the zombie population dies off after initial infection as the US human population continues to grow towards K. The second column resulted in an endemic equilibrium where each population never reaches K, but it never falls to 0. This leads to both species eventually coexisting with one another. The third column is much more chaotic as both populations rise and dip over the years showing a constant struggle for survival.

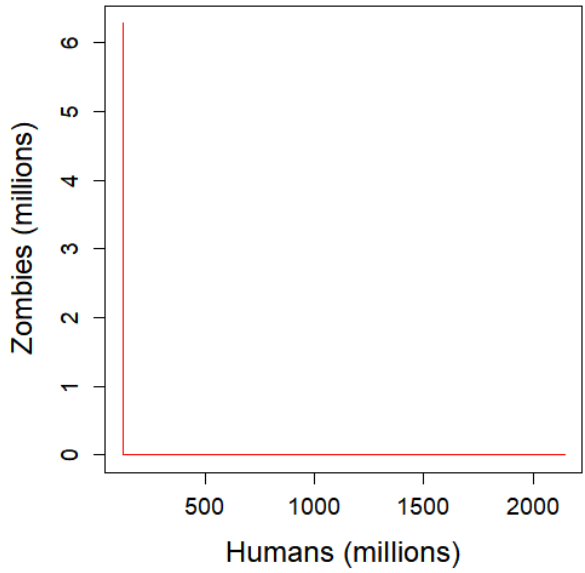
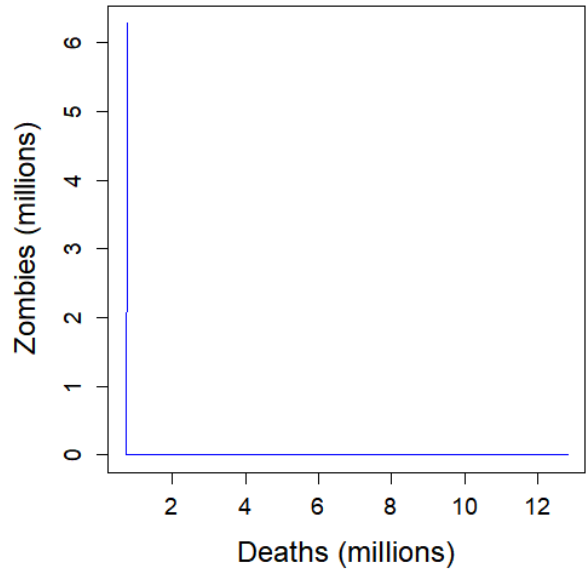
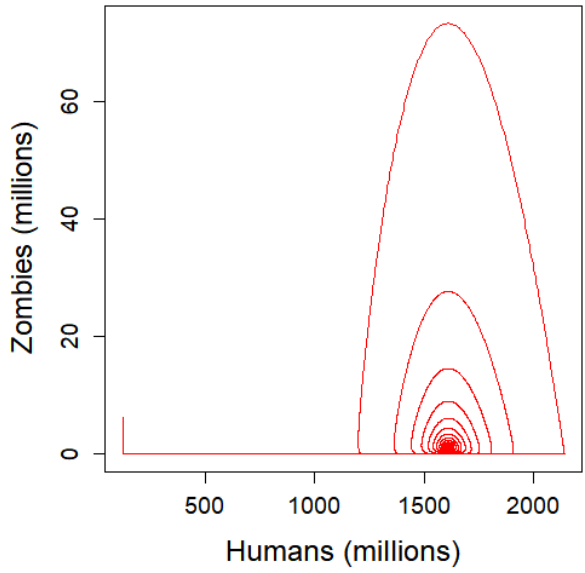
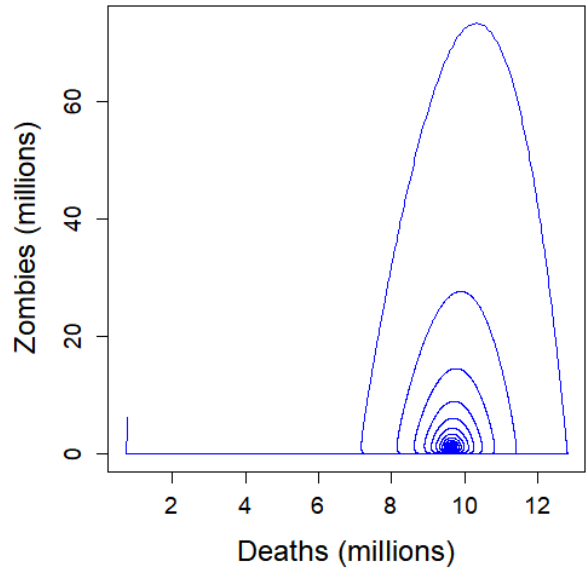

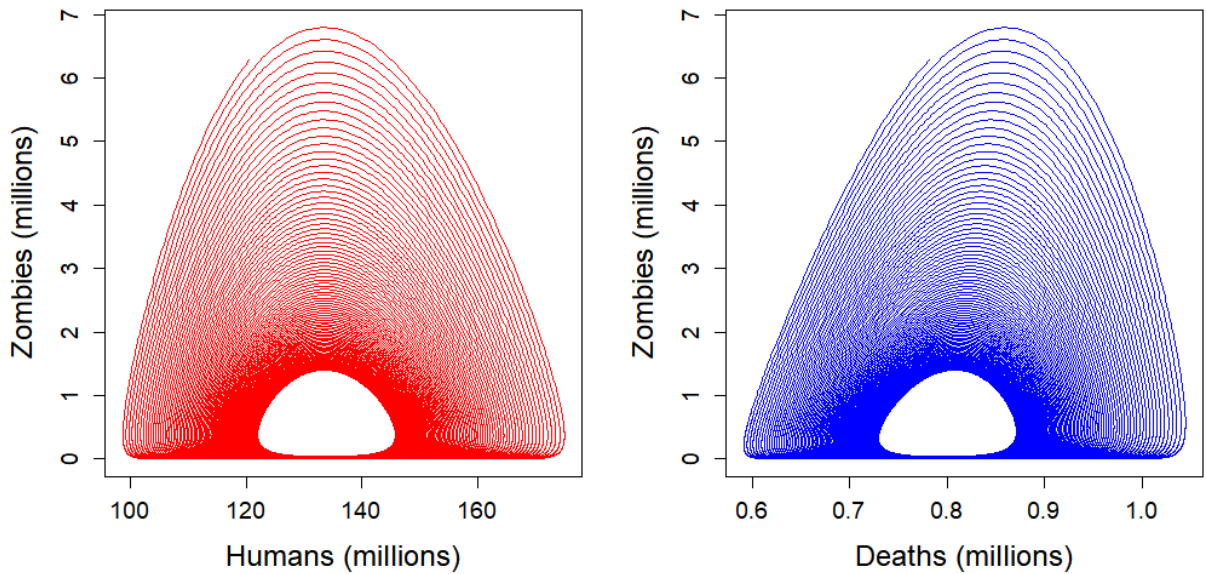

Figure 7. Phase portraits of the extended model. The left, middle, and right columns have β values of 0.015, 0.029, and 0.35, respectively. The top and bottom rows correspond to plots of zombies vs. humans, and deaths vs. zombies, respectively.

## 4. Discussion

We analyzed a mathematical model using stability, evolutionary invasion, and Hopf-bifurcation analyses to determine the long-term prognosis of the United States and the likelihood of a potential zombie uprising. According to our model, the prognosis of the United States remains positive, so long as its birth rate continues to exceed its mortality rate, and no country imports any form of zombie infection. Importantly, our evolutionary invasion analysis shows that an invasion is likely only possible if the ratio of the zombie transmission rate to the zombie death rate is less than the ratio of alive humans to alive and non-decayed dead humans. Unfortunately, if zombies can invade the United States, our stability analysis shows that we would likely have to learn to coexist with zombies, at least until some form of public health intervention is implemented to eradicate them.

According to our results, zombie invasions are theoretically possible, provided a sufficiently large ratio of transmission rate to the zombie death rate. While achieving this ratio is uncommon in nature, a single ant infected with zombie ant fungus can potentially infect entire colonies by seeking elevated locations that promote transmission in tropical climates, such as Brazil, Africa, and Thailand (*Zombie Ants - Real World Sci-Fi Horror Story*, 2020). If a human zombie followed such behavior, this suggests they would seek out a more densely populated area, which would increase the chances of people being infected.

While our work focused on showing the theoretical conditions required for zombies to evolve or invade the United States, there exist many potential future directions. For instance, we could calibrate our model to the transmission cycle of zombie ant fungus and ants to inform the dynamics of zombie evolution. Furthermore, we could also generalize our model to account for additional traits, such as zombie speed or intelligence, or additional zombification stages, such as latent or asymptomatic infection, to gauge their effects on the likelihood of an uprising.

As with all mathematical models, our work has several limitations. To begin, there is a lack of available and reliable data on zombie outbreaks, and our analysis hinged on the functional form of the human mortality rate. Furthermore, research studies on zombie evolution are limited, although recent trends in studying zombie ant fungus are on the rise (whyevolutionistrue, 2020), (Zimmer, 2019). Other important factors from our work include simplifying assumptions on the demographics of zombies and humans alike. Specifically, people with underlying health conditions, disabled people, the elderly, and the young would likely be at high risk of becoming zombies, which could stand to influence the speed that zombism transmits, and its capacity for invasion. Having stated this, the likely advances in science and public health from a zombie outbreak would help to offset such health inequalities, in addition to improving humanity's ability to combat epidemics and eradicate zombies.

Even though zombie ants exist, our main finding indicates human zombies are impossible, from an evolutionary standpoint. Furthermore, upon a situation where human zombies do rise, our work further highlights that the US would likely survive, either by promoting conditions that discourage a zombie invasion or by learning to coexist with zombies in some form of steady-state, at least until the time that medicine or some massive public health intervention turns the tide in humanities' favor.